\documentclass[reqno,12pt,a4paper]{amsart}
\usepackage{graphicx}
\usepackage{psfrag}

\usepackage{mathrsfs}
\usepackage{amssymb}
\usepackage{amsfonts}
\usepackage{latexsym}
\usepackage{amsmath}
\usepackage{amsthm}

\newcommand{\er}[1]{\textrm{(\ref{#1})}}
\def\lb{\label}

\theoremstyle{plain}

\newtheorem{theorem}{\bf Theorem}[section]
\newtheorem{lemma}[theorem]{\bf Lemma}

\newtheorem{corollary}[theorem]{\bf Corollary}
\theoremstyle{remark}

\newcommand{\fr}{\frac}
\newcommand{\tf}{\tfrac}

\renewcommand{\a}{\alpha}           

\renewcommand{\b}{\beta}            \newcommand{\cB}{\mathcal{B}}

\newcommand{\g}{\gamma}           

\newcommand{\G}{\Gamma}

\renewcommand{\l}{\lambda}

\newcommand{\m}{\mu}              

\newcommand{\n}{\nu}

           \newcommand{\cR}{\mathcal{R}}

\renewcommand{\t}{\tau}

\newcommand{\vk}{\varkappa}

  \def\mH{{\mathscr H}}

\def\Z{\mathbb{Z}}
\def\R{\mathbb{R}}
\def\C{\mathbb{C}}
\def\T{\mathbb{T}}
\def\N{\mathbb{N}}

\def\qqq{\qquad}
\def\qq{\quad}

\let\ge\geqslant
\let\le\leqslant

\newcommand{\ca}{\begin{cases}}
\newcommand{\ac}{\end{cases}}
\newcommand{\ma}{\begin{pmatrix}}
\newcommand{\am}{\end{pmatrix}}

\renewcommand{\[}{\begin{equation}}
\renewcommand{\]}{\end{equation}}

\def\wt{\widetilde}
\def\pa{\partial}

\def\no{\noindent}

\def\iy{\infty}

\def\ts{\times}

\def\ss{\subset}

\def\Im{\mathop{\rm Im}\nolimits}

\def\sign{\mathop{\rm sign}\nolimits}

\def\Tr{\mathop{\rm Tr}\nolimits}
\def\const{\mathop{\rm const}\nolimits}

\def\BBox{\hspace{1mm}\vrule height6pt width5.5pt depth0pt \hspace{6pt}}
\def\wh{\widehat}
\def\as{\text{as}}
\def\where{\text{where}}
\def\1{1\!\!1}

\begin{document}

\title[Trace formulas
for fourth order operators] {Trace formulas for fourth order
operators on unit interval,  II}

\date{\today}
\author[Andrey Badanin]{Andrey Badanin}
\address{Mathematical Physics Department, Faculty of Physics, Ulianovskaya 2,
St. Petersburg State University, St. Petersburg, 198904,
 Russia,}
\author[Evgeny Korotyaev]{Evgeny Korotyaev}
\address{
 an.badanin@gmail.com
 \newline
 korotyaev@gmail.com}

\maketitle

\begin{abstract}
We consider self-adjoint fourth order operators on the unit interval
with the Dirichlet type boundary conditions. For such operators we
determine few trace formulas, similar to the case of
Gelfand--Levitan formulas for second order operators.
\end{abstract}

\section {Introduction and main results}
\setcounter{equation}{0}

\subsection{Introduction}
We consider a self-adjoint operator $H$  on $L^2(0,1)$ given by
\[
\lb{op} Hy=(\pa^4+2\pa p\pa +q)y, \qqq {\rm where} \qq
\pa=\fr{d}{dx},
\]
under the Dirichlet type boundary conditions
\[
\lb{4g.dc}
y(0)=y''(0)=y(1)=y''(1)=0.
\]
We assume that the functions $p,q$ are real and satisfy $ p,q\in
L^1(0,1). $ Note that any self-adjoint fourth order operator with
real coefficients may be transformed to the form \er{op}.
It is well known (see, e.g., \cite[Ch.~I.2,~I.4]{N}) that the spectrum  of the operator $H$
consists of real eigenvalues $\m_n,n\in\N$, of multiplicity $\le 2$
labeled by
$$
\m_1\le\m_2\le\m_3\le...,
$$
counted with multiplicities, and they satisfy
$$
\m_n=(\pi n)^4+O(n^{2})\qqq\as\qq n\to\iy.
$$

We recall the famous
Gelfand--Levitan trace formula for a second order operator  $h$
with the Dirichlet boundary conditions on the interval $[0,1]$ given
by
\[
\lb{dc2o}
hy=-y''-py , \qqq \qq y(0)=y(1)=0.
\]
All eigenvalues $\a_n,n\in\N$, of
this operator are simple and labeled
by
$$
\a_{1}<\a_{2}<\a_{3}<...
$$
It is well known (see e.g., \cite{FP} or \er{asaln}), that in the
case $p,p''\in L^1(0,1)$ the eigenvalues $\a_n$ satisfy $ \a_n=(\pi
n)^2-p_0+O(n^{-2})$ as $n\to\iy$, where $p_0=\int_0^1p(t)dt$. In
this case Gelfand and Levitan \cite{GL} determined the following trace
formula
\[
\lb{GLf}
\sum_{n=1}^\iy\big(\a_n-(\pi n)^2+p_0\big)
=\fr{p(0)+p(1)}{4}-\fr{p_0}{2},
\]
where the series converges absolutely.

There are a lot of other results about trace formulas for second
order operators, see the book of Levitan -- Sargsyan \cite{LS} and
references therein. Due to application to the KdV equation on the
circle there are  a lot of papers about an operator  $-\pa^2-p$ on
the circle:  Dubrovin \cite{Du} and Its -- Matveev \cite{IM}
determined trace formulas for so called finite band potentials,
McKean -- van Moerbeke \cite{MvM} and Trubowitz \cite{T} considered the
case of sufficiently smooth potentials; Korotyaev \cite{K}
determined the trace formula for the case $p\in L^2(0,1)$. Note that
the corresponding trace formulas for  the Boussinesq equation were
determined by McKean \cite{M} and for the Camassa -- Holm equation
by Badanin, Klein and Korotyaev \cite{BKK}.

Describe briefly results about trace formulas for fourth and higher
order operators. The trace formulas for operators $(\pa^2+p)^m,p\in
C_\R^\iy[0,1]$ with integer $m\ge 2$ were determined by Gelfand
\cite{G} and Dikii \cite{D1}, \cite{D2}. Sadovnichii \cite{S1},
\cite{S2} obtained trace formulas for even order operators, using
Dikii's approach \cite{D1}, \cite{D2}.   Among  recent papers,
nearest to our subject, we mention results of Badanin -- Korotyaev
\cite{BK5}, G\"ul \cite{Gu}, Nazarov, Stolyarov and Zatitskiy
\cite{NSZ}, see also the review of Sadovnichii -- Podol'skii
\cite{SP} and references therein.

Eigenvalue asymptotics are important for trace formulas: their proof
and the convergence in trace formulas, see e.g., \er{asaln} and
\er{GLf}. Eigenvalue asymptotics for fourth and higher order
operators on the finite interval are much less investigated than for
second order operators.
An operator $\pa^4+\pa p\pa+q$
under the 2-periodic boundary conditions was considered
by Badanin and Korotyaev \cite{BK2}, \cite{BK4}
(for the simpler case $\pa^{4}+q$ see \cite{BK1}).
The sharp eigenvalue asymptotics for the operator $H$
in the class of complex coefficients was determined in \cite{BK6}.
Moreover, there was determined the eigenvalue asymptotics for the
Euler-Bernoulli operator $b^{-1}(af'')'',a,b>0$, on the unit
interval. This operator describes the
bending vibrations of thin beams and plates.
Eigenvalue asymptotics for an operator
$\pa^{2n}+q$ was determined by Akhmerova \cite{Ah},
Mikhailets and Molyboga \cite{MM}. Badanin and
Korotyaev \cite{BK3} determined the eigenvalue asymptotics for a
general case of $2n$ order operators
under the 2-periodic boundary conditions.

We discuss some other results for fourth and higher order operators.
Numerous results about higher order operators with different types
of boundary conditions are expounded in the books of Atkinson
\cite{At} and Naimark \cite{N}. Many papers are devoted to the
inverse spectral problems for these operators, see Barcilon
\cite{B}, Caudill, Perry and Schueller \cite{CPS}, Hoppe, Laptev and
\"Ostensson \cite{HLO}, McLaughlin \cite{Mc}, Papanicolaou
\cite{P}, Yurko \cite{Yu} and so on.

The main goal of the present paper is to determine trace formulas
for fourth order operators on the unit interval.
In fact, our paper is a second part of our previous results from
\cite{BK5}, where we determined trace formulas for fourth
order operators on the circle.

There is a significant difference between trace formulas for
fourth order operators and  for second order operators.
Indeed, we have two coefficients $p,q$ in the fourth order operator,
which corresponds to perturbation by second order operators.
It is possible to determine the following trace formulas:

1) for fix $p$ in terms of $q$,

2) for fix $q$ in terms of $p$,

3) in terms of $p,q$.

\no The case 1) is simpler,
since a perturbation is a function $q$.
The perturbation in the cases 2) and 3) is a second order
operator and it is stronger, than in the case 1).
Therefore, the cases  2) and 3) are more difficult,
than the case 1).

\subsection{Perturbations by second order operators.}
We determine trace formulas for the cases 2) and 3), where
perturbations of the fourth order operators are
second order operators.

Introduce the Sobolev spaces $\mH_m$ and $\mH_m^{0}$, $m\ge 0$, by
$$
\mH_m=\Big\{f\in L^1(0,1): f^{(m)}\in L^1(0,1)\Big\}, \qqq
 \mH_m^0 =\Big\{f\in \mH_m:\int_0^1f(t)dt=0\Big\}.
$$
Recall the following results from \cite{BK6}.
 Let $(p,q)\in \mH_3\ts\mH_1$. Then the eigenvalues $\m_n$ of the
operator $H$ satisfy the asymptotics
\[
\lb{4g.asDir}
\m_n=((\pi n)^2-p_0)^2-\fr{P+p_0^2}{2}+q_0-\wh
V_{cn}+O(n^{-2}),
\]
uniformly on any bounded subset of $(p,q)\in\mH_3\ts \mH_1$,
where
\[
\lb{defp1} f_0=\int_0^1f(t)dt,\qqq
P=\int_0^1(p''(t)+p^2(t))dt,
\]
\[
\lb{V1} \wh V_{cn}=\int_0^1 V(t)\cos(2\pi nt)dt\qqq\forall\qq
n\in\Z,\qqq V=q-\fr{p''}{2}.
\]

Now we formulate our main results on the trace formulas for the
operator $H=\pa^4+2\pa p\pa+q$.
The perturbation in these formulas is the second
order operator $2\pa p\pa+q$, depending on two functions
$p$ and $q$.

\begin{theorem}
\lb{ThTrF}
Let $(p,q)\in \mH_4\ts \mH_2^0 $ and let $\m_n,n\in\N$,
be eigenvalues of the operator $H$. Then the following trace formula
holds true:
\[
\lb{4g.trf3}
\sum_{n\ge1}\big(\m_{n}-((\pi n)^2-p_0)^2+\tf{1}{2}(P+p_0^2)\big)
=-\tf{1}{4}(P-p_0^2+V(0)+V(1)),
\]
where the series converge absolutely and uniformly on any bounded
subset of $\mH_4\ts \mH_2^0 $.

In particular,
if $p=p_0=\const$, then
\[
\lb{trS}
\sum_{n=1}^\iy\big(\m_n-(\pi n)^4+2p_0(\pi n)^2\big)
=-\tf{1}{4}(q(0)+q(1)),
\]
if $q=0$, then
\[
\lb{4g.trq=0}
\sum_{n\ge1}\big(\m_{n}-((\pi n)^2-p_0)^2+\tf{1}{2}(P+p_0^2)\big)
=-\tf{1}{4}(P-p_0^2)+\tf{1}{8}(p''(0)+p''(1)).
\]
\end{theorem}

\no {\bf Remark.}
In the proof of Theorem \ref{ThTrF} we use the presentation
$H=h^2+q-p''-p^2$, where $h$ is the unperturbed  operator,  given by \er{dc2o}. The proof follows our approach from \cite{BK5} and is based on the asymptotic analysis
of the difference of the resolvents of the perturbed operator $H$
and the unperturbed operator $h^2$.

\medskip

Theorem \ref{ThTrF} implies a trace formula for
the operator $h^2$.
This trace formula is known due
to Dikii--Gelfand \cite{D1}, \cite{D2},
\cite{G}. In the following corollary
we extend the Dikii--Gelfand
trace formula
onto a larger class of coefficients $p$.

\begin{corollary}
\lb{corTrh2}
Let $p\in \mH_{4}$ and let $\a_n,n\in\N$, be eigenvalues
of the operator $h$. Then the following trace formula holds true:
\[
\lb{S01}
\sum_{n\ge1}\Big(\a_{n}^2-\big((\pi n)^2-p_0\big)^2-\fr{P-p_0^2}{2}\Big)
=\fr{P+p_0^2}{4}-\fr{p^2(0)+p^2(1)}{4}-\fr{p''(0)+p''(1)}{8},
\]
where the series converges absolutely and uniformly on any bounded
subset of $\mH_4$.
\end{corollary}

\no {\bf Remark.}
1) For the class $p\in C^\iy[0,1]$ the trace formula \er{S01}
was determined by Dikii  \cite{D1},  \cite{D2} and
Gelfand \cite{G}.
Dikii \cite{D1} determined this formula
without any additional restrictions.
Unfortunately, the
results from \cite{D1} contain some mistakes (see \cite[Remark~5 in
Sect~4]{FP} and our discussion in Section 3).
In the second paper \cite{D2} Dikii  determined trace formula \er{S01}
under additional conditions $p^{(2j-1)}(0)=p^{(2j-1)}(1)=0$ for all
$j\in\N$. The proof of Dikii \cite{D1}, \cite{D2} uses the analysis of
the zeta function of the operator.

Gelfand \cite{G} determined the trace formula \er{S01}
under stronger conditions: $p=0$ in some
neighborhoods of the points $0$ and $1$. His proof  is based on the
analysis of an expansion of a trace of the resolvent
in powers of the spectral parameter.

2) There is an open problem to give a transparent proof
of the Dikii--Gelfand trace formulas for the operators $h^m, m\ge 2$,
and, more widely, for the polynomials of $h$.
Corollary \ref{corTrh2} makes only the first step in this direction.

\medskip

Using the trace formula \er{4g.trf3} we can recover
the coefficient of the operator $H$ by the other coefficients
and the spectrum.
Recall that there are trace formulas for second order operators,
see \cite{Du} for finite band potentials, \cite{T},
\cite{L} for smooth potentials,
\cite{K} for potentials from $L^2(0,1)$.
Let $p$ be an 1-periodic smooth function and $p_0=0$.
For any $\t\in\T,\T=\R/\Z$, consider the
shifted operator $h_\t=-\pa^2-p(\cdot+\t)$ on the interval $(0,1)$
with the Dirichlet boundary conditions $y(0)=y(1)=0$.
Let $\a_n(\t),n\in\N$, be eigenvalues of the operator $h_\t$
labeled by $\a_1(\t)<\a_2(\t)<...$
Each function $\a_n(\t),n\in\N$, is 1-periodic and smooth.
The Gelfand--Levitan trace formula \er{GLf} gives
\[
\lb{shGLf}
\sum_{n=1}^\iy\big(\a_n(\t)-(\pi n)^2\big)
=\tf{1}{2}p(\t)\qqq\forall\qq\t\in\T.
\]
If we know $\a_n(\t)$ for all $(n,\t)\in\N\ts\T$,
then we can recover $p$.

For the second order operator $h_\t$ the functions
$\a_n(\t),n\in\N$, satisfy so called Dubrovin system of differential
equations  \cite{Du} with the initial conditions
(so called spectral data)
$$
\textstyle S_n=(\a_n(0), \sign \fr{d}{d\t}\a_n(0)),\qq n\ge 1,
$$ (see p. 324 in \cite{T}).
For given spectral data $S$,
the Dubrovin system has the unique 1-periodic solution
$\a_n(\t)$ for all $(\t,n)\in\T\ts \N$.
Thus, all $\a_n(\t)$ may be determine by $S$.
If we know the  spectral data  $S$,
then using the Dubrovin equation  we recover $\a_n(\t)$,
and then, using the trace formula \er{shGLf}
we recover $p(\t),\t\in\T$.
The open problem is to extend these results onto fourth order
operators.

Consider the fourth order operators.
Introduce the Sobolev spaces $\mH_{m,per}$ and $\mH_{m,per}^0$,
$m\ge 0$ of periodic functions, by
$$
\mH_{m,per}=\Big\{f\in L^1(\T): f^{(m)}\in L^1(\T)\Big\}, \ \
\mH_{m,per}^0=\Big\{f\in\mH_{m,per}:\int_0^1f(t)dt=0\Big\},\ \
\T=\R/\Z.
$$
Let $(p,q)\in\mH_{4,per}\ts\mH_{2,per}^0$.
For any $\t\in\T$ we
define the shifted operator
$H_\t$ on $L^2(0,1)$ by
$$
H_\t=\pa^4+\pa p(x+\t)\pa+q(x+\t)
$$
with the boundary conditions \er{4g.dc}.
Let $\m_n(\t),n\in\N$, be eigenvalues of the shifted operator $H_\t$
labeled by $\m_1(\t)\le\m_2(\t)\le...$, counted with multiplicities.
We formulate our trace formula
\er{ipr1} for the operator $H_\t$.
This formula determines the function
$V(\t),\t\in\T$, by $\m_n(\t)$. Unfortunately,
analysis of the corresponding Dubrovin equations for fourth
order operators is still not carried out. The problem is that
the eigenvalues can have multiplicity 2.

\begin{corollary}
\lb{CorIp1}
Let $(p,q)\in\mH_{4,per}\ts\mH_{2,per}^0$ and let $\m_n(\t),n\in\N$,
be eigenvalues of the operator $H_\t$. Then
there exists $N=N(p,q)\in\N$ such that the functions
$\sum_{n=1}^N\m_{n}(\t)$ and each
$\m_n(\t),n>N$, belong to the space $ C^1(\T)$. Moreover,
they satisfy
\[
\lb{ipr1}
\sum_{n\ge1}\Big(\m_{n}(\t)-\big((\pi n)^2-p_0\big)^2
+\tf{1}{2}(P+p_0^2)\Big)
=-\tf{1}{4}\big(P-p_0^2+2V(\t)\big)\qqq\forall\qq\t\in\T,
\]
where the series converges absolutely and uniformly on $\t\in\T$
and $P=\int_0^1p^2(t)dt$.

In particular, assume that we know $\m_{n}(\t)$
for all $(n,\t)\in \N\ts\T$.
Then

a) If we know $p$, then we can recover $q$.

b) If we know $q,p_0$ and $\int_0^1p^2dt$, then we can recover $p$.
\end{corollary}

\no {\bf Remark.}  Asymptotics \er{4g.asDir} shows that
each eigenvalue $\m_n(\t)$ with $n$ large enough is simple,
and then it is a smooth function of $\t\in\T$.
The situation with other eigenvalues
is complicated:
we don't know how they depend on $\t$.
However, due to Rouch\'e's theorem, we can control their sum,
then it is smooth.

\subsection{Perturbations by functions.}
Now we determine trace formulas for the simplest case:
the perturbation by $q$ for fixed $p$.
In fact, we consider a more general situation,
perturbations of the
operator $H$, given by \er{op}, \er{4g.dc}, by  functions $Q\in \mH_2$.
Let  $\l_n,n\in\N$, be eigenvalues of $H+Q$ labeled by $
\l_1\le\l_2\le\l_3\le..., $ counted with multiplicity.

\begin{theorem}
\lb{ThTrF1} Let $(p,q,Q)\in \mH_4\ts \mH_2\ts \mH_2 $. Let $\m_n$
and $\l_n,n\in\N$, be eigenvalues of the operator $H$ and $H+Q$,
respectively. Then the following trace formula holds true:
\[
\lb{Tr3} \sum_{n\ge1}(\l_{n}-\m_n)=-\tf{1}{4}(Q(0)+Q(1)-2Q_0),
\]
where the series converges absolutely and uniformly on any bounded
subset of $\mH_4\ts \mH_2\ts \mH_2 $.
\end{theorem}

\no {\bf Remark.}
1) Sadovnichii \cite{S1} determined the trace
formula \er{Tr3} for the simplest case $\pa^4+Q$, where
$
Q\in C^\iy[0,1]$ and $Q^{(2j-1)}(0)=Q^{(2j-1)}(1)=0 \
\forall \ j\in\N
$
(see also Remark in Section~4).
Nazarov, Stolyarov and Zatitskiy \cite{NSZ} extended
the results of Sadovnichii onto the larger class
of higher order operators
(see Remark in Section 2).

2) Note that the perturbations by second order operators
in Theorem~\ref{ThTrF} is stronger, than perturbations
by functions in Theorem~\ref{ThTrF1}.
Therefore, we need to analyze more terms of
the perturbation series in the proof of Theorem~\ref{ThTrF},
than in the
proof of Theorem~\ref{ThTrF1}.

\medskip

Now we apply Theorem \ref{ThTrF1} to the case $H=h^2$,
where the operator $h$ is given by \er{dc2o}.
Let $\n_n,n\in\N$, be eigenvalues of the operator $h^2+Q$
labeled by $\n_1\le\n_2\le...$ counting with multiplicities.

\begin{corollary}
\lb{cor}
i) Let $(p,Q)\in \mH_4\ts \mH_2 $. Let $\n_n$ and
$\a_n,n\in\N$, be eigenvalues of the operators $h^2+Q$ and $h$
respectively. Then the following trace formula holds true:
\[
\lb{cor1}
\sum_{n\ge1}(\n_{n}-Q_0-\a_n^2)=-\tf{1}{4}\big(Q(0)+Q(1)-2Q_0\big),
\]
where the series converges absolutely and uniformly on any bounded
subset of $\mH_4\ts \mH_2 $.

ii) Let $(p,Q)\in\mH_4\ts\mH_{2,per}^0 $.
Let $\n_n(\t),n\in\N,\t\in\T$,
be eigenvalues of the operators $h^2+Q(\cdot+\t)$,
labeled by $\n_1(\t)\le\n_2(\t)\le...$, counted with multiplicities.
Then there exists $N=N(Q)\in\N$ such that the functions
$\sum_{n=1}^N\n_{n}(\t)$ and each
$\n_n(\t),n>N$, belong to the space $ C^1(\T)$. Moreover,
they satisfy
\[
\lb{ip2}
\sum_{n\ge1}\big(\n_{n}(\t)-\a_n^2\big)= -\tf{1}{2}Q(\t)\qqq
\forall\qq \t\in\T,
\]
where the series converges absolutely and uniformly on $\t\in\T$.

In particular, if we know $\a_n,\n_{n}(\t)$
for all $(n,\t)\in \N\ts [0,1]$,
then we can recover $Q$.

\end{corollary}

\no {\bf Remark.} Assume that we know $\a_n, \n_{n}(\t)$
for all $(n,\t)\in \N\ts [0,1]$. Then we recover the coefficient $Q$.
Remark that, in the second order case all eigenvalues $\a_{n},n\in\N$ do not determine $p$,  since we need so-called norming constants, e.g. \cite{LS}, \cite{T}. Thus, in order to recover $Q$ we don't need to know $p$, it is sufficiently to know all $\a_n, n\ge 1$.
Of course,
the coefficient $p$ determines all eigenvalues $\a_{n},n\in\N$, and then we recover $Q$.


\medskip

The plan of the paper is as follows.
In Section~2 we consider the perturbation by the function
and prove Theorem \ref{ThTrF1}.
In Section~3  we determine the trace formula for the operator $h^2$.
In fact, this is a simplified version
of the trace formula for the operator $H$.
Some technical proofs are replaced from Section~3 into Section~5.
In Section~4 we determine the trace formula for the operator $H$
and prove Theorem~\ref{ThTrF}.

\section {The proof of Theorem \ref{ThTrF1}}
\setcounter{equation}{0}

 Let $\cB_1$, $\cB_2$ be the sets of all trace class and
Hilbert-Schmidt class operators on $L^2(0,1)$ equipped with the norms
$\|\cdot\|_1, \ \|\cdot\|_2$, respectively. Let
$
(p,q,Q)\in\mH_2\ts\mH_0\ts\mH_0.
$
In order to prove  Theorem \ref{ThTrF1} we need to study  the
resolvents defined by
\[
\lb{resolv}
R_1(\l)=\big(H+Q-\l\big)^{-1},\qq
R(\l)=\big(H-\l\big)^{-1},\qq \cR(\l)=(h^2-\l)^{-1},\qq
\cR_0(\l)=\big(h_0^2-\l\big)^{-1},
\]
where $h_0=-\pa^2$ is equal to $h$ at $p=0$. It is well known (see
\cite[(4.21)]{FP}), that in the case $p,p''\in L^1(0,1)$ the
eigenvalues $\a_n$ satisfy
\[
\lb{asaln} \a_n=(\pi n)^2-p_0+\fr{P-p_0^2}{(2\pi n)^2}+\fr{O(1)}{n^4},
\]
\[
\lb{asaln2} \a_n^2=((\pi n)^2-p_0)^2+
\fr{P-p_0^2}{2}+\fr{O(1)}{n^2}
\]
as $n\to\iy$. Due to asymptotics \er{asaln}, \er{asaln2}, all
resolvents  satisfy
$$
 R_1(\l), R(\l),\cR(\l),   h_0\cR_0(\l) \in \cB_1
$$
 on the corresponding resolvent sets. A proof of the following
results repeats the arguments for the periodic case, see
\cite[Lemma~2.1]{BK5}.

\begin{lemma}
\lb{TL1} Let $(p,q,Q)\in \mH_4\ts\mH_2\ts\mH_2$. Then the following
asymptotics hold true:
\[
\lb{res11}
\|\cR_0(\l)\|_2+\|\cR(\l)\|_2+\|R(\l)\|_2+\|R_1(\l)\|_2=O(k^{-3}),
\]
\[
\lb{res12}
\|h_0\cR_0(\l)\|_2=O(k^{-1}),
\]
\[
\lb{Tr100} \oint_{\G_k}
\Tr\big(\cR_0(\l)(h_0p+ph_0)\cR_0(\l)q\big)d\l=o(1),
\]
as  integer $k\to \iy$, uniformly  on the contours $\G_k$ given by
$$
\G_k=\textstyle\big\{\l\in \C:|\l|^\fr{1}{4}
=\pi\big(k+\fr{1}{2}\big)\big\}.
$$
\end{lemma}

In the Hilbert space $L^2(0,1)$ we introduce the scalar product $
(f,g)=\int_0^1f(x)g(x)dx, $ the norm $\|f\|^2=(f,f)$ and the Fourier
coefficients
$$
f_0=\int_0^1f(x)dx,\qqq\wh f_{cn}=(f,\cos 2\pi nx),\qqq
\wh f_{sn}=(f,\sin 2\pi nx),\qqq n\in\N.
$$
Below we need the following simple result.

\begin{lemma}
Let $Q\in \mH_2^0 ,k\in\N$. Then the following
identity holds true:
\[
\lb{iFV}
\lim_{k\to\iy} \fr{1}{2\pi i}\oint_{\G_k}\Tr Q\cR_0(\l)d\l
=\fr{Q(0)+Q(1)}{4}.
\]
\end{lemma}

\no {\bf Proof.}
We have the Fourier series
\[
\lb{Fs} Q(x)=2\sum_{n=1}^\iy \Big(\wh Q_{cn}\cos 2\pi nx+ \wh
Q_{sn}\sin 2\pi nx\Big).
\]
Let $s_n=\sqrt2\sin\pi nx, n\in\N$. Then
$$
\begin{aligned}
\Tr Q\cR_0(\l)=\sum_{n=1}^\iy \fr{(Q s_n,s_n) }{(\pi n)^4-\l}
=\sum_{n=1}^\iy\fr{\wh Q_{cn}}{\l-(\pi n)^4},
\end{aligned}
$$
since $ (Q s_n,s_n)=-\wh Q_{cn},\qq n\ge 1.
$
This implies
\[
\lb{Trf}
\fr{1}{2\pi i}\oint_{\G_k}\Tr Q\cR_0(\l)d\l
=\sum_{n=1}^k \wh Q_{cn} \qqq
\]
for all $k\ge 1$. Then \er{Fs} and $Q\in \mH_2^0 $ yield \er{iFV}.
$\BBox$

\medskip

Introduce the norm
$$
\|f\|_\iy=\sup_{x\in[0,1]}|f(x)|.
$$

\medskip

\no {\bf Proof of Theorem \ref{ThTrF1}.}
The series
$
\sum_{n\ge1}(\l_{n}-\m_n)
$
converges absolutely and uniformly in $(p,q,Q)$
due to asymptotics \er{4g.asDir}.
We have for integer $k$:
\[
\lb{Tr21}
\sum_{n\ge1}(\l_{n}-\m_n)
=-\fr{1}{2\pi i}\lim_{k\to\iy}\oint_{\G_k}\l\Tr\big(R_1(\l)-R(\l)\big)d\l.
\]
The resolvent identity gives
\[
\lb{Tr22} R_1-R=-RQR+R_1QRQR.
\]
Estimates \er{res11} imply
$$
\big|\Tr \big(R_1(\l)QR(\l)QR(\l)\big)\big|\le
\|R_1(\l)\|_2\|R(\l)\|_2^2\|Q\|_\iy^2=O(k^{-9})
$$
uniformly on $\G_k$,
which yields
\[
\lb{Tr23}
\oint_{\G_k} \l\Tr\big(R_1(\l)QR(\l)QR(\l)\big)d\l=O(k^{-1})
\qqq\as\qq k\to\iy.
\]
Substituting \er{Tr22} into \er{Tr21}
and using \er{Tr23} we obtain
$$
\sum_{n\ge1}(\l_{n}-\m_n)
=\fr{1}{2\pi i}\lim_{k\to\iy}\oint_{\G_k} \l\Tr QR^2(\l) d\l.
$$
Since $R^2(\l)=R'(\l)$, the integration by parts gives
\[
\lb{Tr25}
\sum_{n\ge1}(\l_{n}-\m_n)
=\fr{1}{2\pi i}\lim_{k\to\iy}\oint_{\G_k} \l\Tr QR'(\l) d\l
=-\fr{1}{2\pi i}\lim_{k\to\iy}\oint_{\G_k}\Tr QR(\l) d\l.
\]

The resolvent identity together with
$H=h^2-v,v=p''+p^2-q$ implies
\[
\lb{Tr26} R(\l)=\cR(\l)+R(\l)v\cR(\l).
\]
Estimates \er{res11} give
$$
\big|\Tr \big(R(\l)v\cR(\l)Q\big)\big|\le
\|R(\l)\|_2\|\cR(\l)\|_2\|v\|_\iy\|Q\|_\iy=O(k^{-6})
$$
uniformly on $\G_k$,
which yields
\[
\lb{Tr27} \oint_{\G_k}\Tr \big(R(\l)v\cR(\l)Q\big)d\l=O(k^{-2})
\qqq\as\qq k\to\iy.
\]
Substituting \er{Tr26} into \er{Tr25}
and using \er{Tr27} we obtain
\[
\lb{Tr30}
\sum_{n\ge1}(\l_{n}-\m_n)
=-\fr{1}{2\pi i}\lim_{k\to\iy}\oint_{\G_k}\Tr Q\cR(\l) d\l.
\]

The resolvent identity together with
$h^2=(h_0-p)^2=h_0^2-h_0p-ph_0+p^2$ implies that
the resolvents $\cR,\cR_0$, given by \er{resolv},
satisfy
\[
\lb{Tr29}
\cR=\cR_0-\cR_0A\cR_0+\cR A\cR_0A\cR_0,
\]
where $A=-h_0p-ph_0+p^2$. Let $k\to\iy$.
Estimates \er{res11} give
$$
\big|\Tr \big(\cR_0(\l)p^2\cR_0(\l)Q\big)\big|\le
\|\cR_0(\l)\|_2^2\|p\|_\iy^2\|Q\|_\iy=O(k^{-6})
$$
uniformly on $\G_k$. This
asymptotics and asymptotics \er{Tr100} yield
\[
\lb{res18}
\oint_{\G_k}\Tr \big(\cR_0(\l)A\cR_0(\l)Q\big)d\l
=\oint_{\G_k}\Tr \big(\cR_0(\l)(-h_0p-ph_0+p^2)\cR_0(\l)Q\big)d\l
=o(1).
\]
Estimates \er{res11}, \er{res12} give
$$
\|A\cR_0(\l)\|_2\le 2\|p\|_\iy\|h_0\cR_0(\l)\|_2
+\|p^2\|_\iy\|\cR_0(\l)\|_2
=O(k^{-1}),
$$
then
$$
\big|\Tr \big(\cR(\l) A\cR_0(\l)A\cR_0(\l)Q\big)\big|
\le \|\cR(\l)\|_2\|A\cR_0(\l)\|_2^2\|Q\|_\iy=O(k^{-5})
$$
uniformly on $\G_k$,
which yields
\[
\lb{Tr28}
\oint_{\G_k}\Tr \big(\cR(\l) A\cR_0(\l)A\cR_0(\l)Q\big)d\l=O(k^{-1}).
\]
Substituting \er{Tr29} into \er{Tr30}
and using  \er{res18},  \er{Tr28}
we obtain
$$
\sum_{n\ge1}(\l_{n}-\m_n)
=-\fr{1}{2\pi i}\lim_{k\to\iy}\oint_{\G_k}
\Tr Q\cR_0(\l) d\l.
$$
Identity \er{iFV} implies the trace formula \er{Tr3}
for $Q\in\mH_2^0$.
The trace formula for $Q\in\mH_2$ follows.
$\BBox$

\medskip

\no {\bf Remark.} Nazarov, Stolyarov and Zatitskiy \cite{NSZ}
determined some trace formulas for the operator $\wt H+Q$, where $\wt
H$ is a higher order operator with complex coefficients on the unit
interval and $Q\in L^1(0,1)$ is a complex function.
Authors of \cite{NSZ} determined
$\lim_{N\to\iy}\sum_{n=1}^N(\l_n-\l_n^0-q_0)=X$,
where $\l_n^0,\l_n,n\in\N$, are eigenvalues
of the operators $\wt H,\wt
H+Q$ respectively, and $X$ is expressed
in terms of $q$ and boundary
conditions, the explicit expression is rather
complicated. Note that there are no any control
on a type of convergence
of the series $s$.

\medskip

\no {\bf Proof of Corollary \ref{cor} i).}
Let $H=h^2$. Then
$\l_n=\n_n,\m_n=\a_n^2$ and the trace formula \er{Tr3} gives
\er{cor1}.
\BBox

\section{Diki\"{\i}--Gelfand trace formula}
\setcounter{equation}{0}

Our proof of Theorem \ref{ThTrF} is based on the identity
$H=h^2+q-p''-p^2$. In Theorem \ref{Trh2} we will determine
the trace formula for
the unperturbed operator $h^2=(\pa^2+p)^2.$
In fact, the result of Theorem \ref{Trh2} extends
the result of Corollary \ref{corTrh2}
onto the larger class $p\in\mH_3$.
Our proof is based on the identity
$$
h^2=(\pa^2+p)^2=\pa^4+2\pa p\pa+p''+p^2.
$$
We analyze asymptotics
of the difference of the resolvents of the perturbed
operator $h^2$
and the unperturbed operator $\pa^4$.
Note that we don't use the results and
methods from \cite{D1}, \cite{D2}, \cite{G} in our proof.

Let $p\in\mH_{2}$. Introduce the resolvents
$$
r(z)=(h-z)^{-1},\qq
r_0(z)=(h_0-z)^{-1}.
$$
 Due to asymptotics
\er{asaln} the resolvents (on the corresponding
resolvent sets) satisfy $ r(z)$, $r_0(z) \in \cB_1. $
Moreover,
\[
\lb{estrr0}
\|r_0(z)\|_2+\|r(z)\|_2=O(k^{-1})
\]
as $k\to\iy$ uniformly  on the contours $\g_k\ss\C, k\in\N$, given by
$$
\g_k=\big\{z\in \C:|z|^\fr{1}{2}=\pi
\big(k+\tf{1}{2}\big)\big\}.
$$

Identity
$$
\Tr\big(r(z)-r_0(z)\big)
=\sum_{n=1}^\iy\Big(\fr{1}{\a_n-z}-\fr{1}{(\pi n)^2-z}\Big)
$$
yields
\[
\lb{r1}
-\fr{1}{2\pi i}\int_{\g_k}z^2\Tr\big(r(z)-r_0(z)\big)dz=
\sum_{n=1}^k\Big(\a_n^2-(\pi n)^4\Big)
\]
for all $k\in\N$ large enough.
The following two lemmas give asymptotics of the integral in \er{r1}.

\begin{lemma}
Let $p\in \mH_{2},k\in\N$. Then the following asymptotics holds
true:
\[
\lb{rr3}
-\fr{1}{2\pi i}\int_{\g_k}z^2\Tr\big(r(z)-r_0(z)\big)dz
=\sum_{j=1}^5J_j(k)+O(k^{-1}),
\]
as $k\to\iy$ uniformly on any bounded subset of $\mH_2$, where
\[
\lb{mJj}
J_j(k)=\fr{1}{\pi ij}\int\limits_{\g_k}z\Tr(r_0(z)p)^{j}dz
\qqq\forall \qq j\in\N.
\]

\end{lemma}

\no {\bf Proof.}
Estimates \er{estrr0} give
$$
|\Tr (r_0(z)p)^6r(z)|\le
\|p\|_\iy^6\|r_0(z)\|_2^6\|r(z)\|_2=O(k^{-7})
$$
as $k\to\iy$ on all contours.
Using the resolvent identity
$$
r(z)-r_0(z)=\sum_{j=1}^5(r_0(z)p)^jr_0(z)
+(r_0(z)p)^6r(z)
$$
we obtain \er{rr3},
where
$$
J_j(k)=-\fr{1}{2\pi i}\int_{\g_k}z^2
\Tr(r_0(z)p)^jr_0(z)dz.
$$
Identities
$$
\Tr(r_0(z)p)^jr_0(z)
=\Tr p(r_0(z)p)^{j-1}r_0^2(z)
=\Tr p(r_0(z)p)^{j-1}r_0'(z)
=\fr{1}{j}\Tr\big((r_0(z)p)^{j}\big)'
$$
imply
$$
J_j(k)=-\fr{1}{2\pi ij}\int\limits_{\g_k}z^2
\Tr\big((r_0(z)p)^{j}\big)'dz.
$$
Integration by parts gives \er{mJj}.
$\BBox$

\medskip

 The technical proof of the following lemma see in Section 4.

\begin{lemma}
\lb{LmAsJ} Let $p\in \mH_{2},k\in\N$. Then the sequences
$J_j(k),j=1,...,5$ satisfy
\[
\lb{rr4}
J_1(k)=-\sum_{n=1}^k\Big(2 p_0(\pi n)^2-\fr{p'(1)-p'(0)}{2}\Big)
-\fr{1}{2}\sum_{n=1}^k(\wh {p''})_{cn},
\]
\[
\lb{rr5}
J_2(k)=k\fr{\|p\|^2+p_0^2}{2}-\fr{\|p\|^2-p_0^2}{4}
-\sum_{n=1}^k(\wh{p^2})_{cn}+O(k^{-1}),
\]
\[
\lb{rr7}
J_3(k)=O(k^{-\fr{2}{3}}),\qqq
J_4(k)=O(k^{-\fr{4}{5}}),\qqq
J_5(k)=O(k^{-1})
\]
as $k\to\iy$ uniformly on any bounded subset of $\mH_2$.
\end{lemma}

We prove the Dikii--Gelfand trace formula in our class
of coefficients.

\begin{theorem}
\lb{Trh2}
Let $p\in \mH_{3}$ and let $\a_n,n\in\N$, be eigenvalues
of the operator $h$. Then the trace formula \er{S01} holds true,
where the series converges absolutely and uniformly on any bounded
subset of $\mH_3$.
\end{theorem}

\no {\bf Proof.}
Asymptotics \er{asaln2} shows that the series in \er{S01}
converges absolutely and uniformly on any bounded
subset of $\mH_3$.
Substituting \er{rr4}, \er{rr5}, \er{rr7}
into \er{rr3} we obtain
$$
\begin{aligned}
-\fr{1}{2\pi i}\int_{\g_k}z^2\Tr(r(z)-r_0(z))dz
\\
=-\sum_{n=1}^k\Big(2 p_0(\pi n)^2-\fr{P+p_0^2}{2}\Big)
-\fr{\|p\|^2-p_0^2}{4}
-\fr{1}{2}\sum_{n=1}^k(\wh {p''})_{cn}
-\sum_{n=1}^k(\wh{p^2})_{cn}
+O(k^{-\fr{4}{5}})
\end{aligned}
$$
as $k\to\iy$. This asymptotics and \er{r1} imply
\[
\lb{fs3}
\sum_{n=1}^k\Big(\a_n^2-(\pi n)^4
+2 p_0(\pi n)^2-\fr{P+p_0^2}{2}\Big)
=-\fr{\|p\|^2-p_0^2}{4}
-\fr{1}{2}\sum_{n=1}^k(\wh {p''})_{cn}
-\sum_{n=1}^k(\wh{p^2})_{cn}
+O(k^{-\fr{4}{5}}).
\]
Using the Fourier series for $p\in\mH_3$
$$
\begin{aligned}
p^2(x)=\|p\|^2+2\sum_{n=1}^\iy
\Big((\wh{p^2})_{cn}\cos 2\pi nx+(\wh{p^2})_{sn}\sin 2\pi nx\Big),
\\
p''(x)=p'(1)-p'(0)+2\sum_{n=1}^\iy
\big((\wh{p''})_{cn}\cos 2\pi nx+(\wh{p''})_{sn}\sin 2\pi nx\big),
\end{aligned}
$$
we obtain
\[
\lb{fs1}
\sum_{n=1}^\iy(\wh{p^2})_{cn}=\fr{p^2(0)+p^2(1)}{4}-\fr{\|p\|^2}{2},
\qqq
\sum_{n=1}^\iy(\wh{p''})_{cn}=\fr{p''(0)+p''(1)}{4}-\fr{p(1)-p'(0)}{2}.
\]
Substituting identities \er{fs1} into \er{fs3}
we obtain \er{S01}.
$\BBox$

\medskip

\no {\bf Remark.}
Let $p\in C^\iy[0,1]$.
Dikii \cite[p.~189-190]{D1} determined
the following asymptotics
\[
\lb{asD}
\a_n=(\pi n)^2-p_0+\fr{\wt P}{(2\pi n)^2}+...,
\]
and trace formula
\[
\lb{TrfD1}
\sum_{n=1}^\iy\Big(\a_n^2-(\pi n)^4+2p_0(\pi n)^2
-\fr{\wt P+2p_0^2}{2}\Big)
=\fr{\wt P+2p_0^2}{4}+\fr{p''(0)+p''(1)}{8}-\fr{p^2(0)+p^2(1)}{4},
\]
where
\[
\lb{asq_1D}
\wt P=\|p\|^2-4p_0^2+\fr{1}{3}\big(p'(1)-p'(0)\big).
\]
Asymptotics \er{asD} is in a disagreement with \er{asaln}.
The coefficients
$4$ and $\tf{1}{3}$ in \er{asq_1D} are mistaken,
see also \cite[Remark 4.5]{FP}.

Assume, in addition, that $p_0=0,p^{(2j-1)}(0)=p^{(2j-1)}(1)=0$
for all $j\in\N$.
Then $\wt P=\|p\|^2$ and \er{TrfD1} gives
\[
\lb{asq_1D3}
\sum_{n=1}^\iy\Big(\a_n^2-(\pi n)^4-\fr{\|p\|^2}{2}\Big)
=\fr{\|p\|^2}{4}+\fr{p''(0)+p''(1)}{8}-\fr{p^2(0)+p^2(1)}{4}.
\]
On the other hand, for this case
Dikii \cite[id.~(6.4))]{D2} determined the following trace formula
\[
\lb{asq_1D2}
\sum_{n=1}^\iy\Big(\a_n^2-(\pi n)^4-\fr{\|p\|^2}{2}\Big)
=\fr{\|p\|^2}{4}-\fr{p''(0)+p''(1)}{8}-\fr{p^2(0)+p^2(1)}{4}.
\]
The last two identities are in a disagreement:
the signs before $\tf{1}{8}(p''(0)+p''(1))$
are not coincide.
Theorem \ref{Trh2} shows that the signs
in \er{TrfD1}, \er{asq_1D3} are incorrect.
Thus, trace formula \er{TrfD1} contains two mistakes:
the coefficients in $\wt P$ and the sign before
$\tf{1}{8}(p''(0)+p''(1))$
are incorrect, whereas trace formula \er{asq_1D2}
is valid.

\section{The proof of Theorem~\ref{ThTrF}}
\setcounter{equation}{0}


We prove the main result of our paper.

\medskip

\no {\bf Proof of Theorem \ref{ThTrF}.}
We have the identity
\[
\lb{V}
H=h^2+Q,\qqq
Q=q-p''-p^2 .
\]
The trace formula \er{cor1} and the identity $Q_0=-P$ give
\[
\lb{res171}
\sum_{n\ge1}\big(\m_{n}+P-\a_{n}^2\big)
=-\fr{Q(0)+Q(1)+2P}{4}.
\]

Introduce the sums $S=S(p,q),S_0=S_0(p)$ given by
\[
\lb{trfSS0}
\begin{aligned}
S=\sum_{n\ge1}\Big(\m_{n}-((\pi n)^2-p_0)^2+\fr{P+p_0^2}{2}\Big),
\\
S_0=\sum_{n\ge1}\Big(\a_{n}^2-((\pi n)^2-p_0)^2-\fr{P-p_0^2}{2}
\Big),
\end{aligned}
\]
where the series converges absolutely and uniformly
on  $p,q$ due to asymptotics \er{4g.asDir}, \er{asaln2}.
The trace formula \er{S01} gives
\[
\lb{S012}
S_0
=\fr{P+p_0^2}{4}-\fr{p^2(0)+p^2(1)}{4}-\fr{p''(0)+p''(1)}{8}.
\]
The trace formula \er{res171} and the definitions \er{trfSS0} imply
\[
\lb{res17}
S-S_0=\sum_{n\ge1}\big(\m_{n}+P-\a_{n}^2\big)
=-\fr{Q(0)+Q(1)+2P}{4}.
\]
Substituting identities \er{V}, \er{S012} into \er{res17} we get
$$
S=-\fr{P-p_0^2}{4}+\fr{p''(0)+p''(1)}{8}-
\fr{q(0)+q(1)}{4},
$$
which yields \er{4g.trf3}.
$\BBox$

\medskip

\no {\bf Remark.}
Sadovnichii \cite[p.~308-309]{S1} considered the operator
$H=\pa^4+2\pa p\pa +q$,
where
\[
\lb{Sc}
p,q\in C^\iy[0,1],\qqq
p^{(j)}(0)=p^{(j)}(1)=q^{(2j-1)}(0)=q^{(2j-1)}(1)=0\qqq
\forall\qq j\in\N,
\]
and wrote (without a proof) the following asymptotics
\[
\lb{asS}
\m_n=(\pi n)^4-2p_0(\pi n)^2+q_0+\fr{c_1}{n^2}+\fr{c_2}{n^4}+...
\]
as $n\to\iy$, and trace formula
\[
\lb{TrS}
\sum_{n=1}^\iy\Big(\m_n-(\pi n)^4+2p_0(\pi n)^2-q_0\Big)
=\fr{q_0}{2}-\fr{q(0)+q(1)}{4},
\]
where $c_1,c_2$ are some undetermined constants.
Assume that the operator $H$ has the form
$H=(-\pa^2-p)^2$, where $p$  satisfies \er{Sc}.
Then $q=p''+p^2$ and $q$ satisfies \er{Sc}.
In this case asymptotics  \er{asS}
gives
\[
\lb{asS1}
\m_n=(\pi n)^4-2p_0(\pi n)^2+\|p\|^2+\fr{c_1}{n^2}+\fr{c_2}{n^4}+...
\]
On the other hand, asymptotics
\er{asaln2} yields
\[
\lb{asFP1}
\m_n=\a_n^2=(\pi n)^4-2p_0(\pi n)^2+\fr{\|p\|^2+p_0^2}{2}+\fr{O(1)}{n^2}.
\]
The third term in asymptotics \er{asS1} is in a disagreement with
the corresponding term in \er{asFP1}. Therefore, the term $q_0$ in
\er{asS} is incorrect. Then trace formula \er{TrS}
is not correct also.
Theorem \ref{ThTrF} shows that if $p,q$ satisfy \er{Sc} and $q_0=0$,
then the correct trace formula has the form
$$
\sum_{n=1}^\iy\Big(\m_n-(\pi n)^4+2p_0(\pi n)^2+\fr{\|p\|^2-p_0^2}{2}\Big)
=-\fr{\|p\|^2-p_0^2}{4}-\fr{q(0)+q(1)}{4}.
$$

\medskip

The trace formula \er{S01} is proved in Theorem \ref{Trh2}.
Now we deduce this result immediately from Theorem~\ref{ThTrF}.

\medskip

\no{\bf Proof of Corollary \ref{corTrh2}.} Put
$q=p''+p^2-P\in\mH_2^0$ in Theorem~\ref{ThTrF},
then $H=h^2-P$ and $\m_n=\a_n^2-P$.
Substituting these identities into
\er{4g.trf3} we obtain \er{S01}.
\BBox

\medskip

In order to prove Corollaries \ref{CorIp1}, \ref{cor} ii)
we need the following results.

\begin{lemma}
\lb{LmRTh}
Let $(p,q)\in \mH_2\ts \mH_0^0$. Then
there exists $N=N(\|p\|_{\mH_2},\|q\|_{\mH_0})$
such that
the operator $H$ has exactly $N$ eigenvalues,
counting with multiplicities, in the disc
$\{|\l|<\pi^4(N+\tf{1}{2})^4\}$ and for each integer $n>N$ it has
exactly one simple eigenvalue in the domain
$\{|\l^{1/4}-\pi n|<\tf{\pi}{4}\}$.
There are no other eigenvalues.
\end{lemma}

\no {\bf Proof.} The proof repeats the arguments
from the proof of Lemma 2.5 in \cite{BK6}.
\BBox

\medskip

\no {\bf Proof of Corollary \ref{CorIp1}.}
Let  $(p,q)\in\mH_{4,per}\ts\mH_{2,per}$.
Introduce the function $F(\t)=\sum_{n=1}^N\m_n(\t),\t\in\T$,
where $N$ is given in Lemma \ref{LmRTh},
and the resolvents $R_\t(\l)=(H_\t-\l)^{-1}$.
Then we have
$$
\begin{aligned}
F(\t_1)-F(\t_2)=\fr{1}{2\pi i}\oint_{\G_N}\l
\Tr\big(R_{\t_1}(\l)-R_{\t_2}(\l)\big)d\l,
\\
\m_n(\t_1)-\m_n(\t_2)=\fr{1}{2\pi i}\oint_{\ell_n}\l
\Tr\big(R_{\t_1}(\l)-R_{\t_2}(\l)\big)d\l\qqq\forall\qq n>N,
\end{aligned}
$$
where the contours $\G_N$ and $\ell_n$ are given by
$$
\G_N=\big\{\l\in \C:|\l|^\fr{1}{4}=\pi\big(N+\tf{1}{2}\big)\big\},\qq
\ell_n=\big\{\l\in\C:\{|\l^\fr{1}{4}-\pi n|=\tf{\pi}{4}\}\big\}.
$$
Using the identities
$$
R_{\t_1}(\l)-R_{\t_2}(\l)=
R_{\t_1}(\l)\big(H_{\t_2}-H_{\t_1}\big)R_{\t_2}(\l)
=R_{\t_1}(\l)\big(\pa(p_{\t_2}-p_{\t_1})\pa
+q_{\t_2}-q_{\t_1}\big)R_{\t_2}(\l),
$$
where $f_\t=f(\cdot+\t)$,
we obtain
$$
\big|F^{(k)}(\t_1)-F^{(k)}(\t_2)\big|\le
C\big(\|p_{\t_2}^{(k+1)}-p_{\t_1}^{(k+1)}\|_\iy
+\|p_{\t_2}^{(k)}-p_{\t_1}^{(k)}\|_\iy
+\|q_{\t_2}^{(k)}-q_{\t_1}^{(k)}\|_\iy\big)
$$
for some constant $C>0$ and $k=0,1$, where
$f^{(0)}=f,f^{(k)}=\fr{d^kf}{d\t^k}$.
These estimates imply $F\in C^1(\T)$.
The similar arguments show that
$\m_n\in C^1(\T)$ for all $n>N$.
The trace formula \er{4g.trf3} yields \er{ipr1}.
\BBox

\medskip

\no {\bf Proof of Corollary \ref{cor} ii).}
Repeating the arguments from the proof of Corollary \ref{CorIp1}
we deduce that  the functions
$\sum_{n=1}^N\n_{n}(\t)$ and
$\n_n(\t),n>N$, belong to $ C^1(\T)$.
The trace formula \er{cor1} yields \er{ip2}.
$\BBox$

\section{Asymptotics of $J_j$}
\setcounter{equation}{0}

In this Section we will consider the integrals $J_j,j=1,...,5$ given
by
\[
\lb{mJj1}
J_j(k)=\fr{1}{\pi ij}\int\limits_{\g_k}z\Tr(r_0(z)p)^{j}dz,\qqq
k\in\N,
\]
see \er{mJj}, and prove Lemma \ref{LmAsJ}.
Now we will determine asymptotics of the sequence $J_1$.
Introduce the functions
$$
s_n=\sqrt 2\sin \pi n x,\qqq c_n=\sqrt 2\cos \pi n x,
\qqq n\in\N.
$$

\medskip

\no {\bf Proof of identity \er{rr4} in Lemma \ref{LmAsJ}.}
Substituting the identity
$$
\Tr pr_0(z)
=\sum_{n=1}^\iy \fr{(ps_n,s_n)}{(\pi n)^2-z},
$$
into \er{mJj1} we obtain
$$
J_1(k) =\fr{1}{\pi i}\int_{\g_k}z\Tr p\big(r_0(z)\big)dz
=\fr{1}{\pi i}\int_{\g_k}z\sum_{n=1}^\iy
\fr{(ps_n,s_n)}{(\pi n)^2-z}dz
=-2 \sum_{n=1}^k(\pi n)^2(ps_n,s_n).
$$
Using the identities
$$
(ps_n,s_n)=p_0-\wh p_{cn} =p_0-\fr{p'(1)-p'(0)-(\wh
{p''})_{cn}}{(2\pi n)^2},
$$
we obtain \er{rr4}.
$\BBox$

\medskip

Introduce the coefficients
$$
\vk_n=(p,c_n),\qqq n\in\N.
$$
Note that
\[
\lb{psq}
(ps_n,s_m)=\fr{\vk_{n-m}-\vk_{m+n}}{\sqrt2}\qqq\forall\qq m,n\in\N.
\]
The integration by parts gives
$$
\vk_n=(p,c_n)=\fr{1}{(\pi n)^2}\Big(\sqrt 2 \big(p'(0)-(-1)^np'(1)\big)+(p'',c_n)\Big)
\qqq\forall\qq n\in\N,
$$
which yields the estimate
\[
\lb{estq}
|\vk_n|\le \fr{C}{n^2}\qqq\forall\qq n\in\N,\qqq\where\qq
C=\fr{1}{\pi^2}\Big(\sqrt 2 \big(|p'(0)|+|p'(1)|\big)
+\max_{n\in\N}|(p'',c_n)|\Big).
\]
Now we determine asymptotics of the sequences $J_j(k),j=3,4,5$.

\medskip

\no {\bf Proof of asymptotics \er{rr7} in Lemma \ref{LmAsJ}.}
Let $k\in\N$.
Identity \er{mJj1} gives
\[
\lb{J3AA}
J_3=\fr{1}{3\pi i}\int_{\g_k}z\Tr(r_0(z)p)^{3}dz=\fr{A_1+A_2}{3\pi i},
\]
where
$$
A_1=\int_{\g_k}z\Tr\big(pr_0(z)(r_0(z)p)^2\big)dz,\qqq
A_2=-\int_{\g_k}z\Tr\big([p,r_0(z)](r_0(z)p)^2\big)dz,
$$
$[a,b]=ab-ba$.
Moreover,
$$
\begin{aligned}
A_2=-\int_{\g_k}z\Tr\big(r_0(z)[p,h_0]r_0^2(z)pr_0(z)p\big)dz
=-\int_{\g_k}z\Tr\big([p,h_0]r_0'(z)(pr_0(z))^2\big)dz
\\
=-\fr{1}{3}\int\limits_{\g_k}z\Tr\big([p,h_0]r_0(z)(pr_0(z))^2\big)'dz
=\fr{1}{3}\int\limits_{\g_k}\Tr\big([p,h_0]r_0(z)(pr_0(z))^2\big)dz,
\end{aligned}
$$
where we have used the integration by parts.
Let $k\to\iy$. The identity
$[p,h_0]=-2p'\pa-p''$ and the estimate
$$
\big|\Tr\big(p''r_0(z)(pr_0(z))^2\big)\big|\le\|p''\|_\iy\|p\|_\iy^2\|r_0(z)\|_2^3
=O(k^{-3})
$$
uniformly on $\g_k$, give
\[
\lb{asA2p}
A_2=-\fr{2}{3}\int\limits_{\g_k}\Tr\big(r_0(z)(pr_0(z))^2p'\pa\big)dz
+O(k^{-1}).
\]
Using the estimate
$
\|\pa r_0(z)\|_2=O(1)
$
uniformly on $\g_k$ we obtain
\[
\lb{est1a}
\big|\Tr\big(r_0(z)(pr_0(z))^2p'\pa\big)\big|
\le\|\pa r_0(z)\|_2\|p\|_\iy^2\|p'\|_\iy\|r_0(z)\|_2^2
=O(k^{-2})
\]
uniformly on $\g_k$.
Using the estimate $\|r_0(z)\|\le k^{-\a}$ for $|\Im z|\ge  k^\a,\a>0,$
we obtain
\[
\lb{est2a}
\big|\Tr\big(r_0(z)(pr_0(z))^2p'\pa\big)\big|
\le\|\pa r_0(z)\|_2\|p\|_\iy^2\|p'\|_\iy\|r_0(z)\|^2
=O(k^{-2\a})
\]
as $|\Im z|\ge  k^\a$.
Since the length of part $|\Im z|< k^\a$ of the contour $\g_k$ is
$O(k^\a)$, estimate \er{est1a} shows that
its contribution to the integral \er{asA2p} is
$O(k^{-2+\a})$.
Moreover, estimate \er{est2a} yields that
its contribution to the integral \er{asA2p} is
$O(k^{-2\a+2})$. Let $\a=\tf{4}{3}$. Then \er{asA2p}
gives
\[
\lb{asA2}
A_2=O(k^{-\tf{2}{3}}).
\]

Furthermore,
$$
\begin{aligned}
A_1=\int_{\g_k}z\Tr\big(pr_0^2(z)pr_0(z)p\big)dz
=\int_{\g_k}z\Tr\big(pr_0'(z)pr_0(z)p\big)dz
\\
=\fr{1}{2}\int_{\g_k}z\Big(\Tr\big(pr_0(z)\big)^2p\Big)'dz
=-\fr{1}{2}\int_{\g_k}\Tr\big((pr_0(z))^2p\big)dz,
\end{aligned}
$$
where we have used the integration by parts. Using the identity
$$
\Tr\big((pr_0(z))^2p\big)=\sum_{m,n=1}^\iy \fr{\a_{mn}}{(z-z_m)(z-z_n)}
$$
where
\[
\lb{Amnk}
 z_n=(\pi n)^2,\qqq \a_{mn}=(p^2s_m,s_n)(ps_m,s_n),
\]
we obtain
$$
A_1=-\fr{1}{2}\int_{\g_k}\sum_{m,n=1}^\iy \fr{\a_{mn}}{(z-z_m)(z-z_n)}dz
=\sum_{m=1}^k\sum_{n=k+1}^\iy\fr{\a_{mn}}{z_n-z_m}
=-\fr{1}{2\pi^2}\sum_{m=1}^k\sum_{n=k+1}^\iy\fr{\a_{mn}}{n^2-m^2}.
$$
Estimates
$$
\fr{1}{n^2-m^2}\le \fr{1}{(k+1)^2-k^2}\le\fr{1}{2k}\qqq \forall\qq 1\le m\le k,\ n\ge k+1,
$$
imply
\[
\lb{estJ3}
|A_1|\le\fr{1}{2\pi^2k}\sum_{m=1}^k\sum_{n=k+1}^\iy|\a_{mn}|.
\]
Substituting  \er{psq} into \er{Amnk} we obtain
$$
\a_{mn}=\fr{1}{2}(\wt \vk_{n-m}-\wt \vk_{n+m})(\vk_{n-m}-\vk_{n+m}),
\qqq\where\qq \wt \vk_n=(p^2,c_n),
$$
which yields
$$
\sum_{m=1}^k\sum_{n=k+1}^\iy |\a_{mn}|\le
2\sum_{m,n=1}^\iy|\wt \vk_{m}||\vk_{n}|<\iy.
$$
Then estimate \er{estJ3} gives $A_1=O(k^{-1})$ as $k\to\iy$.
This asymptotics
and \er{asA2} and \er{J3AA} give the first asymptotics in \er{rr7}.

We show  the second asymptotics in \er{rr7}. Identities \er{mJj1}
imply
\[
\lb{est3}
|J_4|\le\fr{1}{4\pi}\int_{\g_k}|z||\Tr(r_0(z)p)^{4}||dz|.
\]
Let $k\to \iy$. Asymptotics \er{estrr0} gives
\[
\lb{est1}
|\Tr(r_0(z)p)^{4}|\le\|r_0(z)\|_2^4\|p\|_\iy^4=O(k^{-4})
\]
on the contours $\g_k$.
Using the estimate $\|r_0(z)\|\le k^{-\b}$ for $|\Im z|\ge  k^\b,\b>0$,
we obtain
\[
\lb{est2}
|\Tr(r_0(z)p)^{4}|\le\|r_0(z)\|^4\|p\|_\iy^4=O(k^{-4\b})
\]
as $|\Im z|\ge  k^\b$.
Since the length of the part $|\Im z|<k^\b$ of the contour $\g_k$ is
$O(k^\b)$, estimate \er{est1} shows that
its contribution to the integral \er{est3} is
$O(k^{-2+\b})$.
Moreover, estimate \er{est2} yields that
the  contribution of the rest part
of the contour is
$O(k^{-4\b+4})$. Let $\b=\tf{6}{5}$. Then \er{est3}
yields the second asymptotics in \er{rr7}.

We prove the third asymptotics in \er{rr7}.
Asymptotics \er{estrr0} gives
$$
|\Tr(r_0(z)p)^{5}|\le\|r_0(z)\|_2^5\|p\|_\iy^5=O(k^{-5})
$$
as $k\to \iy$ on the contours $\g_k$. Then identities \er{mJj1}
imply
$$
|J_5(k)|\le\fr{1}{5\pi}\int_{\g_k}|z||\Tr(r_0(z)p)^{5}||dz|=O(k^{-1})
$$
as $k\to \iy$, which yields the last asymptotics in \er{rr7}.
$\BBox$

\medskip

Introduce the coefficients
\[
\lb{idamn}
a_{mn}=(ps_n,s_m)^2=\fr{(\vk_{n-m}-\vk_{m+n})^2}{2},\qqq m,n\in\N,
\]
where we have used \er{psq}.
In order to determine asymptotics of $J_2(k)$
we need the following preliminary results.

\begin{lemma}
Let $p\in \mH_{3}$. Then

i)  The following identity holds true:
\[
\lb{sr2}
J_2=A_1-A_2,
\]
where
\[
\lb{A1A2}
A_1=k\|p\|^2-\sum_{n=1}^k(\wh{p^2})_{cn},\qqq
A_2=\sum_{m=1}^k\sum_{n=k+1}^\iy a_{mn}\theta_{mn},\qqq
\theta_{mn}=\fr{ n^2+ m^2}{ n^2- m^2}.
\]

ii)  The coefficient $A_2$ satisfies:
\[
\lb{idS12a} A_2=\fr{2k+1}{4}B_0+B_1+B_2+B_3,
\]
where
\[
\lb{S112}
\begin{aligned}
B_0=\sum_{\ell=1}^k\vk_{\ell}^2,\qqq
B_1=\sum_{m=1}^k\sum_{n=m+k+1}^\iy a_{mn}\theta_{mn},
\\
B_2=\fr{1}{2}\sum_{m=1}^k\sum_{n=k+1}^{m+k}
(\vk_{m+n}^2-2\vk_{n-m}\vk_{m+n})
\theta_{mn},
\qqq
B_3=\fr{1}{4}\sum_{\ell=1}^k\vk_{\ell}^2\sum_{n=k+1}^{k+\ell}
\fr{\ell}{2n-\ell}.
\end{aligned}
\]
Moreover,
\[
\lb{aS2}
B_0(k)=\|p\|^2-p_0^2+O(k^{-3}),\qq
B_1(k)=O(k^{-2}),\qq B_2(k)=O(k^{-1}),\qq
B_3(k)=O(k^{-1})
\]
as $k\to\iy$ uniformly on any bounded subset of $\mH_2$.

iii) The sequence $A_2(k)$ satisfies the asymptotics
\[
\lb{sr3}
A_2(k)=(2k+1)\fr{\|p\|^2-p_0^2}{4}+O(k^{-1})
\]
as $k\to\iy$ uniformly on any bounded subset of $\mH_2$.

\end{lemma}

\no {\bf Proof.}
i) Substituting the identity
$$
\Tr\big(r_0(z)p\big)^2
=\sum_{m,n=1}^\iy
\fr{a_{mn}}{(z-z_n)(z-z_m)},\qqq z_n=(\pi n)^2,
$$
into \er{mJj1} we obtain
\[
\lb{sr21}
\begin{aligned}
J_2=\fr{1}{2\pi i}\int_{\g_k}z\Tr(r_0(z)p)^2dz
=\fr{1}{2\pi i}\int_{\g_k}z\sum_{m,n=1}^\iy
\fr{a_{mn}}{(z-z_n)(z-z_m)}dz
\\
=\fr{1}{4\pi i}\int_{\g_k}\sum_{m,n=1}^\iy a_{mn}
\Big(\fr{1}{z-z_n}+\fr{1}{z-z_m}+\fr{z_n+z_m}{(z-z_n)(z-z_m)}\Big)dz
=F_1+F_2+F_3,
\end{aligned}
\]
where
$$
\begin{aligned}
F_1=\sum_{m,n=1}^k a_{mn}+
\fr{1}{2}\sum_{m,n=1:\atop m\ne n}^k a_{mn}(z_n+z_m)\Big(\fr{1}{z_m-z_n}+\fr{1}{z_n-z_m}\big)
=\sum_{m,n=1}^k a_{mn},
\\
F_2=\fr{1}{2}\sum_{m=1}^k\sum_{n=k+1}^\iy a_{mn}
\Big(1+\fr{z_n+z_m}{z_m-z_n}\Big),
\qqq
F_3=\fr{1}{2}\sum_{n=1}^k\sum_{m=k+1}^\iy a_{mn}
\Big(1+\fr{z_n+z_m}{z_n-z_m}\Big)=F_2.
\end{aligned}
$$
Using these identities and  \er{sr21}, we obtain \er{sr2}, where
$A_2$ is defined by the second identity in \er{A1A2} and
$$
A_1=\sum_{m=1}^k\sum_{n=1}^\iy a_{mn}=\sum_{m=1}^k\sum_{n=1}^\iy (ps_m,s_n)^2.
$$
The Parseval identity $\sum_{n=1}^\iy(f,s_n)^2=\|f\|^2$ gives
$$
A_1=\sum_{m=1}^k\|ps_m\|^2
=\sum_{m=1}^k(\|p\|^2-(\wh{p^2})_{cm}),
$$
which shows that $A_1$ satisfies the first identity in \er{A1A2}.

ii) Definition \er{idamn} and the definition of $A_2$
in \er{A1A2} give
\[
\lb{idS12} A_2=\fr{1}{2}\sum_{m=1}^k\sum_{n=k+1}^{m+k} \vk_{n-m}^2
\theta_{mn}+B_1+B_2,
\]
and using the new variable $n-m=\ell$ we obtain
$$
\sum_{m=1}^k\sum_{n=k+1}^{m+k} \vk_{n-m}^2
\theta_{mn}=\sum_{m=1}^k\sum_{\ell=k+1-m}^{k} \vk_{\ell}^2
\fr{(m+\ell)^2+m^2}{ \ell(\ell+2m)}=\sum_{\ell=1}^k \vk_{\ell}^2\sum_{m=k+1-\ell}^{k}
\Big(\fr{m}{\ell}+\fr{m+\ell}{ \ell+2m}\Big).
$$
Due to the identities
$$
\sum_{m=k+1-\ell}^{k}\fr{m}{\ell}=\fr{2k+1-\ell}{2},
\qq
\sum_{m=k+1-\ell}^{k}\fr{m+\ell}{ \ell+2m}=\sum_{n=k+1}^{k+\ell}\fr{n}{2n-\ell}
=\fr{\ell}{2}+\fr{1}{2}\sum_{n=k+1}^{k+\ell}\fr{\ell}{2n-\ell}
$$
we get
\[
\lb{s3w3}
\sum_{m=1}^k\sum_{n=k+1}^{m+k} \vk_{n-m}^2
\theta_{mn}=\fr{2k+1}{2}\sum_{\ell=1}^k \vk_{\ell}^2
+\fr{1}{2}\sum_{\ell=1}^k \vk_{\ell}^2\sum_{n=k+1}^{k+\ell}\fr{\ell}{2n-\ell}.
\]
Substituting this identity into \er{idS12} we obtain \er{idS12a}.

We will prove the first asymptotics in \er{aS2}.
The Parseval identity
$\|f_0\|^2+\sum_{n=1}^\iy(f,c_n)^2=\|f\|^2$ implies
$$
\sum_{\ell=1}^\iy \vk_\ell^2=\|p\|^2-p_0^2.
$$
Estimate \er{estq} gives
$$
\Big|\sum_{\ell=k+1}^\iy \vk_\ell^2\Big|
\le C^2\sum_{\ell=k+1}^\iy \fr{1}{n^4}\le C^2\int_{k}^\iy \fr{dx}{x^4}
=\fr{C^2}{3k^3}.
$$
Then
the definition of $B_0$ in \er{S112} yields
 the first asymptotics in \er{aS2}.

We will prove the second asymptotics in \er{aS2}.
We have the estimates
$$
\theta_{mn}=1+\fr{2m^2}{ n^2- m^2}\le 1+\fr{2k^2}{(m+k+1)^2-m^2}\le 3,
\qqq 1\le m\le k\le n-m-1.
$$
Moreover, definition \er{idamn} implies
$$
a_{mn}\le \vk_{m-n}^2+\vk_{m+n}^2\qqq\forall\qq m,n\in\N.
$$
Then the definition of $B_1$ in \er{S112}
and estimate \er{estq} give
$$
\begin{aligned}
0\le B_1\le3\sum_{m=1}^k\sum_{n=m+k+1}^\iy (\vk_{n-m}^2+\vk_{m+n}^2)
=3\Big(\sum_{m=1}^k\sum_{\ell=k+1}^\iy \vk_{\ell}^2
+\sum_{m=1}^k\sum_{\ell=2m+k+1}^\iy \vk_{\ell}^2\Big)
\\
\le 6\sum_{m=1}^k\sum_{\ell=k+1}^\iy \vk_{\ell}^2=6k\sum_{\ell=k+1}^\iy \vk_{\ell}^2
\le 6kC^2\sum_{\ell=k+1}^\iy \fr{1}{n^4}\le 6kC^2\int_{k}^\iy \fr{dx}{x^4}
=\fr{2C^2}{k^2},
\end{aligned}
$$
which yields
the second asymptotics in \er{aS2}.

We will prove the third asymptotics in \er{aS2}.
The definition of $B_2$ in \er{S112} and estimate \er{estq} imply
$$
\begin{aligned}
|B_2|\le\fr{1}{2}\sum_{m=1}^k\sum_{n=k+1}^{m+k} |\vk_{m+n}|
(|\vk_{m+n}|+2|\vk_{n-m}|)\theta_{mn}
\\
\le\fr{C^2}{2}\sum_{m=1}^k\sum_{n=k+1}^{m+k} \fr{1}{(n-m)^2}
\Big(\fr{1}{(n+m)^2}+\fr{2}{(n-m)^2}\Big)\theta_{mn}.
\end{aligned}
$$
Using the estimates
$$
\begin{aligned}
\fr{1}{(n-m)^2}
\Big(\fr{1}{(n+m)^2}+\fr{2}{(n-m)^2}\Big)\theta_{mn}
=\fr{(3n^2+3m^2+2mn)(n^2+ m^2)}{(n-m)^3(n+m)^5 }
\\
\le\fr{3(n^2+ m^2)^2}{(n-m)^3(n+m)^5 }\le\fr{3}{(n-m)^3(n+m)},
\qqq  1\le m<n,
\end{aligned}
$$
we obtain
$$
\begin{aligned}
|B_2|\le\fr{3C^2}{2}\sum_{m=1}^k\sum_{n=k+1}^{m+k}\fr{1}{(n-m)^3(n+m) }
=\fr{3C^2}{2}\sum_{m=1}^k\sum_{\ell=k+1-m}^{k}\fr{1}{\ell^3(\ell+2m) }
\\
=\fr{3C^2}{2}\sum_{\ell=1}^k\fr{1}{\ell^3}\sum_{m=k+1-\ell}^{k}\fr{1}{\ell+2m}
\le\fr{3C^2}{2(k+2)}\sum_{\ell=1}^k\fr{1}{\ell^3}\sum_{m=k+1-\ell}^{k}1
=\fr{3C^2}{2(k+2)}\sum_{\ell=1}^k\fr{1}{\ell^2},
\end{aligned}
$$
which yields
the third asymptotics in \er{aS2}.

We will prove the fourth asymptotics in \er{aS2}.
The definition of $B_3$ in
\er{S112} and the estimate
$$
\sum_{n=k+1}^{k+\ell}\fr{1}{2n-\ell}\le
\fr{\ell}{k+1}\qqq\forall\qq 1\le \ell\le k,
$$
imply
$$
0\le B_3\le\fr{1}{4(k+1)}\sum_{\ell=1}^k \ell^2\vk_{\ell}^2
\le\fr{C^2}{4(k+1)}\sum_{\ell=1}^k \fr{1}{\ell^2}.
$$
where we have used \er{estq}.
This asymptotics yields
the fourth asymptotics in \er{aS2}.

iii) Asymptotics \er{aS2} and identity \er{idS12a} yield
\er{sr3}.
$\BBox$

\medskip

\no{\bf Remark.} If $p$ is a trigonometric polynomial, then $\vk_n=0$ for all $n\in\N$
large enough, hence  $B_1=B_2=0$ for all $k\in\N$
large enough.

\medskip

\no {\bf Proof of asymptotics \er{rr5} in Lemma \ref{LmAsJ}.}
Substituting the definition of $A_1$ in \er{A1A2}
and identity \er{sr3} into \er{sr2} we obtain
\er{rr5}.
$\BBox$

\medskip

\setlength{\itemsep}{-\parskip} \footnotesize
\no
{\bf Acknowledgments.} {Various parts of this paper were written
during Evgeny Korotyaev's stay in Aarhus
University, Denmark. He is grateful to the institute for the
hospitality. His study was partly supported by the RFFI grant  No 11-01-00458 and
by  project  SPbGU  No 11.38.215.2014.}


\end{document}